To appear in the 9th International Conference on Blockchain Technology and Applications (ICBTA 2026) proceedings

# Peer-Relative Trust-Guided Update Screening for Robust Semi-Decentralized Federated Learning under Label-Flipping Attacks

Ajay Kumar Shrestha[1[0000-0002-1081-7036]] and Sandhya Joshi[2[0000-0002-9289-3787]]

Vancouver Island University, Nanaimo BC V9R 5S5, Canada
[1]ajay.shrestha@viu.ca, [2]sandhya.joshi@viu.ca

**Abstract.** Trust and reputation mechanisms can reduce the influence of malicious clients in federated learning (FL), but strict update filtering is difficult under non-independent and identically distributed (non-IID) data because honest local updates may temporarily reduce performance on a shared validation proxy. Extending an earlier trust-based semi-decentralized FL (SDFL) framework, this paper operationalizes the trust signals and introduces peer-relative update screening that combines a robust validation-utility score with direction- and norm-based update quality. The method is evaluated with 50 MNIST clients, 20% persistent label-flip attackers, Dirichlet heterogeneity levels of 0.5 and 0.1, 100 communication rounds, and five paired random seeds. Under moderate heterogeneity, the proposed method reaches 98.37% final accuracy while rejecting 95.82% of malicious updates and retaining 90.30% of benign updates; the final benign and malicious trust distributions are completely separated (KS=1.00). Under severe heterogeneity, it reaches 95.25% accuracy, rejects 88.55% of malicious updates, and retains 76.90% of benign updates (KS=0.96). Trust weighting alone preserves slightly higher utility in the most heterogeneous condition, but accepts all malicious submissions. The results therefore distinguish influence mitigation from explicit containment and quantify the utility-selectivity trade-off introduced by screening. The study provides an empirically validated off-chain governance core that can feed blockchain-based audit and incentive layers in SDFL.

## 1 Introduction

Federated learning (FL) enables a population of clients to train a shared model while keeping raw training records local, thereby reducing the need to centralize privacy-sensitive data [1]. This advantage does not guarantee trustworthy participation. A malicious or faulty client can manipulate its local data or model updates, and the aggregation process may incorporate those updates without direct access to the underlying records. Trustworthiness surveys therefore identify poisoning resistance, contribution assessment, and robust coordination as persistent challenges for practical FL deployments [2–4].

Classical robust aggregation methods address this problem by limiting the influence of extreme updates. Coordinate-wise median and trimmed-mean procedures provide well-established Byzantine robustness under their modeling assumptions [5]. In FL, however, client data are commonly non-IID, and statistically unusual updates need not be malicious. Recent work consequently combines robustness with client trust or reputation. FLGT uses guiding trust to defend against label-flipping attacks [6]; partially trusted FL uses trusted participants, cosine similarity, and outlier detection [7]; and FLAURA couples adaptive trust evaluation with hard filtering to suppress poison leakage [8]. These studies reinforce the value of trust-aware aggregation, while also highlighting the central difficulty addressed here: differentiating harmful updates from legitimate statistical heterogeneity.

A complementary line of work uses blockchain, smart contracts, and reputation mechanisms to make FL coordination auditable and incentive compatible. Dynamic reputation consensus and adaptive incentives have been proposed for trustworthy and fair FL [9], while surveys of blockchain-based FL emphasize reputation, transparent coordination, and decentralized trust as recurring architectural themes [10]. Earlier semi-decentralized FL research introduced blockchain-assisted trust penalization [11], incentive and penalty mechanisms [12], and, most recently, a conceptual multi-factor trust-based incentive framework with trust decay, recovery, admission control, screening, IPFS-backed artifacts, and smart-contract enforcement [13]. That framework was intentionally theoretical: the trust dimensions were not yet operationalized, and its screening behavior was not evaluated under controlled adversarial and non-IID conditions.

This paper addresses that empirical gap. Rather than attempting to evaluate the entire blockchain and incentive stack at once, it isolates the off-chain trust-governance core that determines which client updates should influence learning. A central design issue is that a strict rule requiring every local update to improve a shared validation proxy is unsuitable for heterogeneous FL: an honest local model can temporarily reduce proxy accuracy while still contributing useful information after aggregation. We therefore define update utility relative to the other updates observed in the same round, using a median/MAD robust score, and combine this signal with directional and norm-based quality assessment. Dynamic trust then modulates aggregation, while the proposed method additionally rejects extreme low-utility or low-quality updates before a trust-weighted robust aggregation step.

The study is organized around three research questions: RQ1 asks whether trust-aware aggregation stabilizes learning under label-flip poisoning compared with FedAvg and trimmed mean; RQ2 asks whether trust-guided screening explicitly contains malicious submissions while retaining benign updates; and RQ3 asks how increasing non-IID heterogeneity changes the utility-containment trade-off.

The contributions are fourfold. First, we operationalize the previously conceptual multi-factor trust model using peer-relative validation utility, temporal consistency, update similarity/quality, and recent participation. Second, we combine the peer-relative utility signal with directional and norm-based update quality in a heterogeneity-aware screening rule and integrate the retained updates with dynamic trust weighting and coordinate-wise robust aggregation. Third, we conduct a paired five-seed evaluation in which all methods share the same data split, client partition, attacker identities, participant schedule, and local shuffle schedule, enabling clean component-level comparisons. Fourth, the results expose an important distinction: trust weighting can mitigate adversarial influence without excluding malicious submissions, whereas screening provides explicit containment at a measurable utility cost when client data become highly heterogeneous.

## 2 Related Work

### 2.1 Poisoning Defense and Robust Aggregation

FedAvg established iterative model averaging as a practical FL baseline [1], but ordinary averaging is vulnerable when adversarial clients can submit corrupted updates. Byzantine-robust distributed learning introduced coordinate-wise median and trimmed mean as robust alternatives with provable statistical properties [5]. In federated settings, however, heterogeneity complicates anomaly detection because benign updates can deviate substantially from the aggregate direction. A recent robust FL algorithm for partially trusted environments uses trusted-client outputs, cosine similarity, and Local Outlier Factor to identify malicious participants [7]. FLGT specifically targets label flipping by constructing a guiding trust signal from last-layer update similarities and uses that signal to identify and rescale suspicious updates [6]. FLAURA further combines adaptive trust evaluation with hard filtering to address poison leakage under non-IID and adaptive attacks [8]. These approaches motivate combining trust with robust aggregation, but they also differ in assumptions about trusted reference data, attack knowledge, and the form of client exclusion.

### 2.2 Reputation, Trust, and Blockchain-Assisted FL

Trust management in FL also extends beyond one-round anomaly scores. Rashid et al. proposed dynamic reputation consensus with temporal discounting, data-quality and reliability factors, adaptive incentives, and blockchain-backed reputation management [9]. Broader surveys of blockchain-based FL identify reputation architectures, smart contracts, tamper-resistant audit trails, and decentralized coordination as recurring trust mechanisms [10]. Previous studies have linked trust penalization, incentives, and blockchain-assisted accountability in semi-decentralized FL [11–13].

The framework [13] defined dynamic multi-factor trust, decay/recovery, participation policies, and blockchain/IPFS enforcement architecture, but did not empirically test the learning-side mechanism. The present paper is therefore not a repetition of that framework. It contributes an operational, peer-relative trust formulation, a revised screening rule appropriate for non-IID data, component-isolated baselines, and controlled poisoning experiments. Blockchain execution, rewards, slashing, and IPFS overhead remain outside the experimental scope here and are discussed only as the downstream governance layer that can consume the evaluated trust decisions.

## 3 Peer-Relative Trust-Guided Learning Core

### 3.1 System Setting

We consider a client population $\mathcal{N} = \{1, \ldots, N\}$ participating in round-based semi-decentralized FL, with N = 50 in the experiments. At communication round t, a method-independent schedule selects a subset $\mathcal{A}_t \subset \mathcal{N}$ of K = 10 clients. The current global model is $\theta_t$. Each selected client $i$ initializes from $\theta_t$, trains for one local epoch on its private partition, obtains local parameters $\theta_{i,t}(local)$, and returns the update $\Delta_{i,t} = \theta_{i,t}(local) - \theta_t$. Raw data never leave the client partition. A fixed coordinator-side validation proxy $\mathcal{V}$ is used only to evaluate submitted updates; $Acc(\theta;\ \mathcal{V})$ denotes the classification accuracy of model $\theta$ on $\mathcal{V}$. The subset that passes screening is denoted by $\mathcal{G}_t \subseteq \mathcal{A}_t$. In an operational semi-decentralized deployment, the same scoring logic may be executed by a rotating or cluster-level coordinator; blockchain-based auditing and incentives can be applied after scoring as proposed in [13].

### 3.2 Peer-Relative Validation Utility

For client $i$ in round $t$, the raw validation effect $g_{i,t}$ is the change in proxy accuracy obtained by applying that client's update to the current global model:

$$g_{i,t} = Acc(\theta_t + \Delta_{i,t}; \mathcal{V}) - Acc(\theta_t; \mathcal{V}). \quad (1)$$

A strict condition $g_{i,t} \geq 0$ is brittle under non-IID data because a locally useful but distribution-specific update can reduce shared-proxy accuracy in isolation. We therefore compare each submitted update with its round peers. Let $m_t$ be the median of the K validation effects. The median absolute deviation (MAD) and robust scale $s_t$ are defined as

$$m_t = median_i(g_{i,t}), \quad MAD_t = median_i|g_{i,t} - m_t|, \quad s_t = 1.4826 \cdot MAD_t.$$

The factor 1.4826 is the normal-consistency correction for MAD, given by $1/\Phi^{-1}(0.75)$, where $\Phi$ is the standard normal cumulative distribution function. If the MAD-based scale is numerically degenerate, the implementation uses $max(\sigma_t, 0.001)$, where $\sigma_t$ is the standard deviation of the K round-level effects. The 0.001 scale floor corresponds to the accuracy resolution of the 1,000-example validation proxy and prevents near-zero denominators. The peer-relative robust utility score is $z_{i,t} = (g_{i,t} - m_t)/s_t$. To obtain a bounded component for trust calculation, $z_{i,t}$ is clipped to [−6, 6] and mapped through the logistic function; $clip(x, a, b) = min(max(x, a), b)$. Clipping to [−6, 6] bounds the logistic-mapped score to approximately [0.0025, 0.9975]. These are numerical design choices rather than empirically optimized values.

$$A_{i,t} = 1/(1 + exp(-clip(z_{i,t}, -6, 6))). \quad (2)$$

Thus, $A_{i,t} = 0.5$ corresponds to the round median. The normalization does not require prior knowledge of which clients are malicious. The robustness of this normalization depends on the round median remaining representative. Although adversaries constitute 20% of the client population, their proportion among selected clients may vary across rounds.

### 3.3 Similarity-Based Update Quality

The quality component $D_{i,t}$ combines update direction and magnitude. A coordinate-wise median of submitted updates forms the robust reference $\Delta\star_t$. Let $c_{i,t} = cos(\Delta_{i,t}, \Delta\star_t)$ denote cosine similarity, and let $r_{i,t} = \|\Delta_{i,t}\|_2 \,/\, median_{j \in \mathcal{A}_t}\|\Delta_{j,t}\|_2$ denote the ratio between the client-update norm and the median submitted-update norm. We define

$$D_{i,t} = ((1 + c_{i,t})/2) \cdot exp(-max(0, r_{i,t} - \kappa)), \quad \kappa = 1.5. \tag{3}$$

The first term maps directional agreement to [0, 1], while the exponential term penalizes unusually large update norms. This component complements validation utility: $A_{i,t}$ measures peer-relative contribution on $\mathcal{V}$, whereas $D_{i,t}$ measures structural agreement with a robust round reference.

### 3.4 Dynamic Trust, Decay, and Recovery

Following the dynamic-trust structure introduced in [13], each client maintains $T_{i,t} \in [0, 1]$. In addition to $A_{i,t}$ and $D_{i,t}$, the model uses temporal consistency $C_{i,t}$ and recent participation $U_{i,t}$. Consistency is an exponentially weighted moving $average$ $(EWMA)$: for an accepted update, $C_{i,t} = (1-\rho)C_{i,t-1} + \rho A_{i,t}$; for a rejected update, $C_{i,t} = (1-\rho)C_{i,t-1}$, with $\rho = 0.20$. $U_{i,t}$ is the fraction of the previous $L$ = 10 rounds in which client $i$ was selected. The base trust score for an active client is

$$B_{i,t} = 0.40A_{i,t} + 0.20C_{i,t} + 0.30D_{i,t} + 0.10U_{i,t}. \tag{4}$$

Let $\bar{A}_{i,t}$ denote the mean of client $i$'s recent A-values over the $L$-round history window. A bounded recovery term rewards improvement relative to that recent history:

$$T_{i,t} = clip(B_{i,t} + \eta(1 - B_{i,t})max(0, A_{i,t} - \bar{A}_{i,t}), 0, 1), \quad \eta = 0.10. \tag{5}$$

Clients begin with $T_{i,0} = 0.60$ and $C_{i,0} = 0.50$. An inactive client receives one exponential decay step per inactive round, $T \leftarrow T \cdot exp(-\lambda)$, with $\lambda = 0.02$. Equivalently, after consecutive inactivity from the last active round $t_0$, $T_{i,t} = T_{i,t0} \cdot exp[-\lambda(t - t_0)]$. The initialization values, 10-round history window, EWMA rate of 0.20, and decay rate of 0.02 serve as fixed reference settings for reproducibility rather than universally optimal parameters.

### 3.5 Screening and Aggregation

The proposed method, denoted C4 in the experimental comparison, accepts client $i$'s update into $\mathcal{G}_t$ only if both peer-relative utility and similarity quality are satisfied:

$$z_{i,t} \geq -3.5 \quad \text{and} \quad D_{i,t} \geq 0.55.$$

The utility threshold rejects only extreme lower-tail validation effects relative to the current round, rather than requiring positive absolute gain. Accepted updates are aggregated with a trust-weighted coordinate-wise trimmed mean. Let $M_{\mathrm{t}} = |\mathcal{G}_{\mathrm{t}}|$ be the number of accepted updates and $q_{\mathrm{t}} = \lfloor 0.10 M_{\mathrm{t}} \rfloor$, where $\lfloor \cdot \rfloor$ denotes the floor operator. For each parameter coordinate, the $q_{\mathrm{t}}$ smallest and $q_{\mathrm{t}}$ largest values are removed; trust weights are carried through the coordinate-wise sort and renormalized over the surviving values. This preserves both robust trimming and dynamic trust weighting. The utility threshold of −3.5, quality threshold of 0.55, and 10% trimming ratio are likewise fixed reference settings; their sensitivity is not evaluated in this study.

# 4 Experimental Methodology

## 4.1 Dataset, Model, and FL Protocol

Experiments use MNIST. For each seed, 1,000 images are deterministically held out from the 60,000-image training split as the coordinator validation proxy $\mathcal{V}$, and the remaining 59,000 training samples are distributed across 50 clients. The test set remains untouched until global evaluation. A compact CNN with two convolutional layers (16 and 32 channels), two max-pooling operations, a 64-unit fully connected layer, and a 10-class output layer is used, totaling 105,866 trainable parameters. Each selected client performs one local stochastic-gradient-descent (SGD) epoch with batch size 32, learning rate 0.01, and momentum 0.9. Ten of the 50 clients participate in each of 100 communication rounds. Test accuracy is recorded in round 1, every five rounds, and the final round.

Two label-skew non-IID settings are generated using Dirichlet partitioning: concentration $\alpha$ = 0.5 represents moderate heterogeneity and $\alpha$ = 0.1 represents severe heterogeneity. The partition generator enforces at least 20 training examples per client. The benign scenario S0 is evaluated at $\alpha$ = 0.5. The poisoning scenario S1 is evaluated at both $\alpha$ = 0.5 and $\alpha$ = 0.1.

## 4.2 Threat Model and Label-Flipping Attack

In S1, 20% of the 50 clients are designated adversarial at the start of each seeded experiment and remain adversarial for the full run. During local training, an attacker maps every class label y to (y + 1) mod 10 before optimization. The attack therefore corrupts the local classification objective while retaining the ordinary local-training pipeline. The coordinator does not know attacker identities; identities are used only for post-hoc measurement of malicious and benign acceptance rates. The study assumes authenticated client identities and focuses on data poisoning rather than Sybil attacks, backdoors, inference attacks, or a malicious coordinator.

The source code, experimental configurations, and supporting artifacts are available in the project repository [14].

### 4.3 Baselines and Fair Comparison Controls

Five methods are evaluated to isolate the contribution of each component of the proposed approach. C0 (FedAvg) uses ordinary mean aggregation without trust, screening, or robust filtering. C1 (Trimmed Mean) replaces mean aggregation with an unweighted coordinate-wise trimmed mean using a 10% trimming ratio. C2 (Static Reputation) assigns aggregation weights using the current round's normalized peer-relative utility score $A_{i,t}$, without temporal trust or update screening. C3 (Dynamic Trust) uses the dynamic trust score $T_{i,t}$ as the aggregation weight but accepts all submitted updates. C4 (Proposed Trust + Screening) combines dynamic trust with peer-relative update screening and trust-weighted coordinate-wise trimmed-mean aggregation. The component-wise differences among the five evaluated methods are summarized in Table 1.

**Table 1.** Evaluated methods and component-wise comparison.

| ID | Method | Trust/ Reputation | Screening | Robust Aggregation |
|---|---|---|---|---|
| C0 | FedAvg | None | No | No |
| C1 | Trimmed Mean | None | No | Yes |
| C2 | Static Reputation | Current $A_{i,t}$ only | No | No |
| C3 | Dynamic Trust | Dynamic $T_{i,t}$ | No | No |
| C4 | Proposed Trust + Screening | Dynamic $T_{i,t}$ | Yes | Yes |

For a fixed seed, C0-C4 use the same train/validation split, client partition, adversarial client identities, participant schedule, and per-client/per-round mini-batch shuffle schedule. This paired design is important because non-IID FL can exhibit large seed-to-seed variance. Participant selection is intentionally independent of trust in this conference study so that differences can be attributed to weighting, screening, and robust aggregation rather than to different client samples. Trust-based admission control remains part of the broader framework [13] but is reserved for future evaluation.

### 4.4 Metrics and Statistical Analysis

We report final test accuracy, normalized area under the learning curve (AULC), benign update acceptance rate, malicious update acceptance rate, and the two-sample Kolmogorov–Smirnov (KS) statistic comparing final benign and malicious reputation or trust scores, as applicable to each method. AULC is the trapezoidal integral of recorded test accuracy normalized by the evaluated round range and captures stability throughout training rather than only the final checkpoint. Each condition is repeated with five $seeds$ $(0-4)$, and results are reported as mean $\pm$ standard deviation (SD). C4 is compared with available baselines using paired seed-wise t-tests and exact Wilcoxon signed-rank tests. With only five paired observations, the Wilcoxon test has coarse p-value resolution; inferential results are therefore interpreted together with paired differences, dispersion, and learning trajectories rather than as stand-alone evidence.

# 5 Results

## 5.1 Benign Learning Performance

Under S0 with $\alpha = 0.5$, all four evaluated methods converge to approximately 98.5% final test accuracy. FedAvg reaches 98.58% $\pm$ 0.14%, dynamic trust 98.54% $\pm$ 0.08%, and C4 98.45% $\pm$ 0.14%. C4 therefore differs from FedAvg by only $-0.126$ percentage points in final accuracy. Its AULC is 94.81%, compared with 95.06% for FedAvg and 95.29% for dynamic trust. The screening mechanism retains 89.90% $\pm$ 2.22% of benign submissions. These results show that peer-relative screening avoids the benign-update rejection collapse that would result from requiring every non-IID local update to produce non-negative absolute validation gain.

## 5.2 Robustness under Label-Flipping Poisoning

The poisoning scenario produces a clear separation between ordinary/robust aggregation and trust-aware weighting. At $\alpha = 0.5$, FedAvg finishes at 72.82% $\pm$ 32.80% and trimmed mean at 77.47% $\pm$ 28.32%, indicating substantial instability across seeds. Static reputation, dynamic trust, and C4 remain stable near 98.4%, with final accuracies of 98.39% $\pm$ 0.06%, 98.41% $\pm$ 0.09%, and 98.37% $\pm$ 0.18%, respectively. The trajectory-level difference is also substantial: C4 AULC is 94.06%, versus 80.67% for FedAvg and 86.08% for trimmed mean. The mean paired AULC improvements are +13.38 and +7.98 percentage points, respectively.

Under severe heterogeneity ($\alpha = 0.1$), all methods face a harder learning problem. C4 reaches 95.25% $\pm$ 1.59%, compared with 79.63% $\pm$ 16.95% for FedAvg and 89.40% $\pm$ 10.35% for trimmed mean. Its AULC is 86.62%, giving paired improvements of +8.16 points over FedAvg and +4.38 points over trimmed mean. Static reputation and dynamic trust reach slightly higher final accuracies (96.54% and 96.43%), which is important for interpreting the role of screening: the main added value of C4 is not a universal accuracy gain over weighting-only methods, but explicit control over which submissions enter aggregation. The corresponding five-seed learning trajectories are shown in Fig. 1.

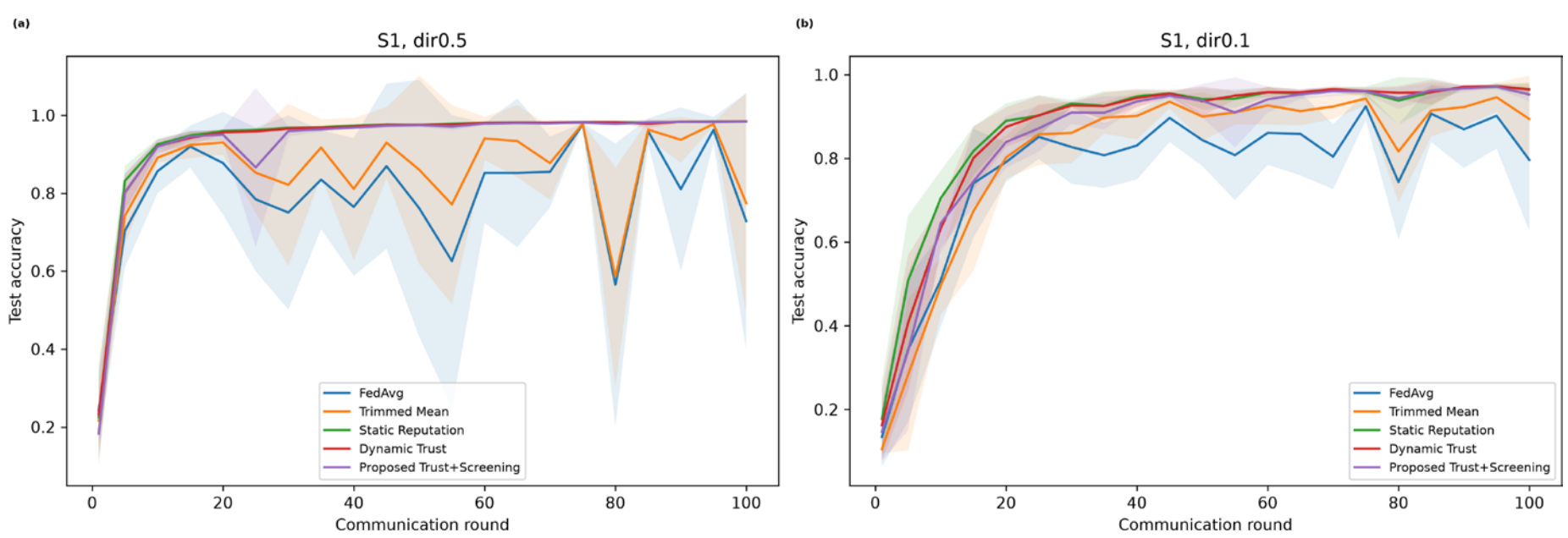

**Fig. 1.** Test accuracy under 20% label-flip poisoning: (a) Dirichlet α = 0.5; (b) Dirichlet α = 0.1. Curves are five-seed means with ±1 SD shading.

The complete numerical results across all evaluated conditions are summarized in Table 2.

**Table 2.** Final five-seed results. Accuracy, AULC, and acceptance rates are proportions; values are mean ± SD.

| Cond. | Method | Final | AULC | Mal. accept. | Ben. accept. | KS |
|---|---|---|---|---|---|---|
| S0, $\alpha$ =.5 | C0 | 0.986 ± 0.001 | 0.951 ± 0.006 | - | 1.000 ± 0.000 | - |
| S0, $\alpha$ =.5 | C1 | 0.985 ± 0.001 | 0.948 ± 0.006 | - | 1.000 ± 0.000 | - |
| S0, $\alpha$ =.5 | C3 | 0.985 ± 0.001 | 0.953 ± 0.004 | - | 1.000 ± 0.000 | - |
| S0, $\alpha$ =.5 | C4 | 0.985 ± 0.001 | 0.948 ± 0.005 | - | 0.899 ± 0.022 | - |
| S1, $\alpha$ =.5 | C0 | 0.728 ± 0.328 | 0.807 ± 0.031 | 1.000 ± 0.000 | 1.000 ± 0.000 | - |
| S1, $\alpha$ =.5 | C1 | 0.775 ± 0.283 | 0.861 ± 0.015 | 1.000 ± 0.000 | 1.000 ± 0.000 | - |
| S1, $\alpha$ =.5 | C2 | 0.984 ± 0.001 | 0.951 ± 0.006 | 1.000 ± 0.000 | 1.000 ± 0.000 | 1.000 ± 0.000 |
| S1, $\alpha$ =.5 | C3 | 0.984 ± 0.001 | 0.948 ± 0.007 | 1.000 ± 0.000 | 1.000 ± 0.000 | 1.000 ± 0.000 |
| S1, $\alpha$ =.5 | C4 | 0.984 ± 0.002 | 0.941 ± 0.014 | 0.042 ± 0.039 | 0.903 ± 0.010 | 1.000 ± 0.000 |
| S1, $\alpha$ =.1 | C0 | 0.796 ± 0.170 | 0.785 ± 0.027 | 1.000 ± 0.000 | 1.000 ± 0.000 | - |
| S1, $\alpha$ =.1 | C1 | 0.894 ± 0.104 | 0.822 ± 0.035 | 1.000 ± 0.000 | 1.000 ± 0.000 | - |
| S1, $\alpha$ =.1 | C2 | 0.965 ± 0.015 | 0.891 ± 0.016 | 1.000 ± 0.000 | 1.000 ± 0.000 | 1.000 ± 0.000 |
| S1, $\alpha$ =.1 | C3 | 0.964 ± 0.014 | 0.882 ± 0.017 | 1.000 ± 0.000 | 1.000 ± 0.000 | 0.955 ± 0.051 |
| S1, $\alpha$ =.1 | C4 | 0.953 ± 0.016 | 0.866 ± 0.023 | 0.115 ± 0.038 | 0.769 ± 0.027 | 0.960 ± 0.055 |

## 5.3 Malicious-Update Containment and Trust Separation

C4 adds an explicit containment capability that is not visible from final accuracy alone. At $\alpha$ = 0.5, only 4.18% ± 3.95% of malicious submissions are accepted, equivalent to rejecting 95.82% of attacker updates, while 90.30% ± 1.02% of benign submissions are retained. At $\alpha$ = 0.1, malicious acceptance rises to 11.45% ± 3.83% and benign retention falls to 76.90% ± 2.66%. The mechanism therefore continues to reject 88.55% of malicious updates under severe heterogeneity, but with a larger false-rejection cost for statistically atypical benign clients.

Trust trajectories independently support the screening result. Under $\alpha$ = 0.5, benign-client trust stabilizes near 0.48 while attacker trust declines toward approximately 0.03 by the end of training; the final empirical distributions are completely separated (mean KS = 1.00). Under $\alpha$ = 0.1, the final KS remains 0.96 ± 0.05 despite wider variability caused by client heterogeneity. The trust score therefore remains discriminative even when the screening boundary becomes less selective. The corresponding benign/malicious update-acceptance rates and trust trajectories under both heterogeneity settings are shown in Fig. 2.

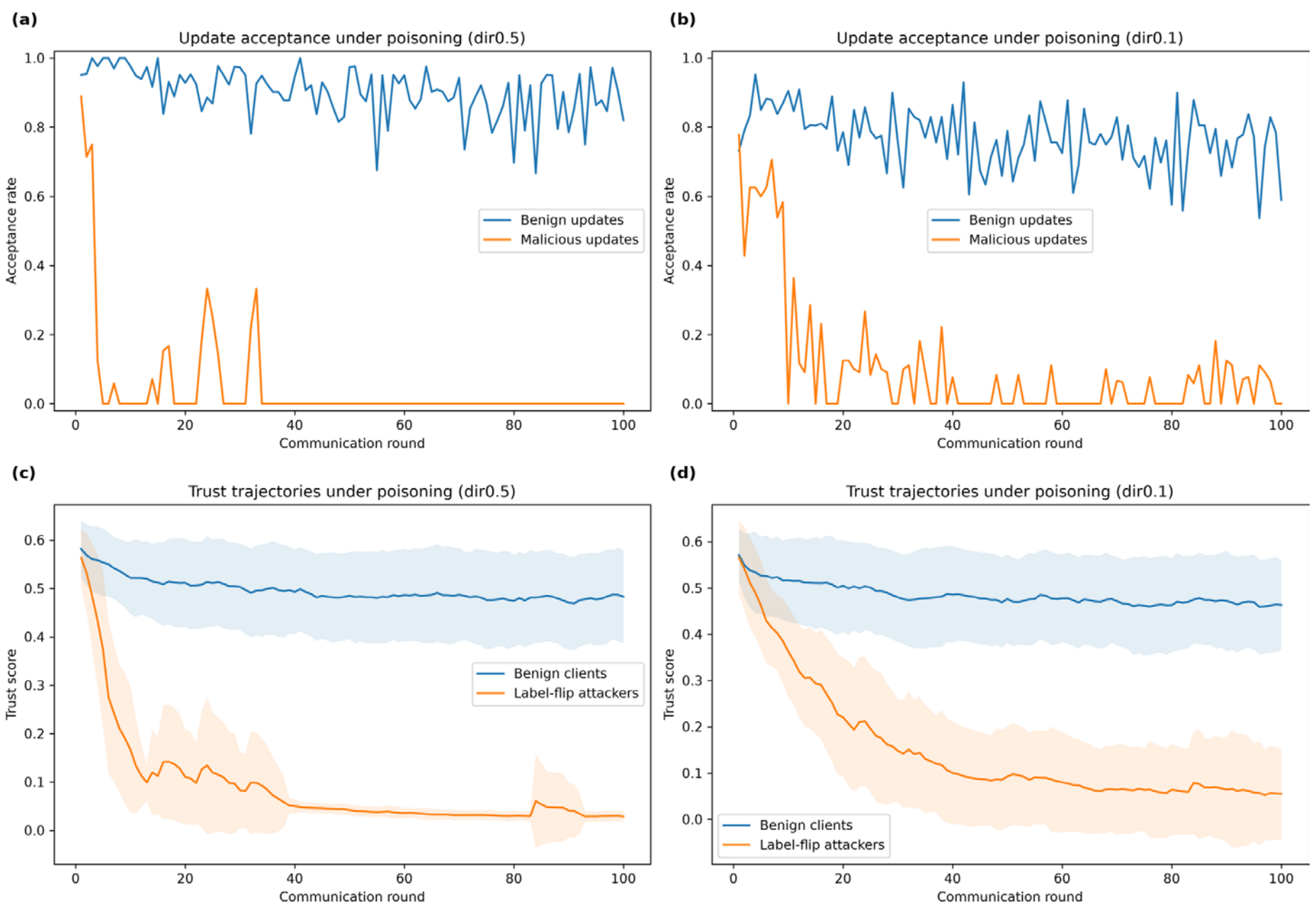

**Fig. 2.** C4 governance signals under poisoning: update acceptance for (a) Dirichlet $\alpha$ = 0.5 and (b) $\alpha$ = 0.1; trust trajectories for (c) $\alpha$ = 0.5 and (d) $\alpha$ = 0.1. Malicious acceptance and trust decline substantially relative to benign clients.

## 5.4 Paired Statistical Comparisons

Because FedAvg and trimmed mean exhibit large final-round variance under poisoning, AULC provides the more stable paired comparison. For $\alpha$ = 0.5, the C4-versus-FedAvg AULC difference is +0.1338 (paired t-test $p$ = 0.0022), and C4-versus-trimmed-mean is +0.0798 ($p$ = 0.0016). For $\alpha$ = 0.1, the corresponding differences are +0.0816 ($p$ < 0.001) and +0.0438 ($p$ = 0.0032). In all four comparisons, the exact two-sided

Wilcoxon $p$-value is 0.0625, reflecting the coarse resolution available with only five paired seeds rather than contradictory directionality. The results are therefore interpreted primarily through paired effect magnitudes, variance, acceptance behavior, and learning trajectories.

# 6 Discussion

## 6.1 Influence Mitigation versus Explicit Containment

A counterintuitive finding is that strong predictive accuracy can coexist with accepting every malicious submission. Both C2 and C3 maintain high model accuracy despite accepting all malicious updates, as utility- or trust-based weighting reduces their influence without explicitly excluding them. C4 adds a different property: the system can explicitly declare a submitted update unacceptable before aggregation. At $\alpha$ = 0.5, this containment is obtained with almost no final-accuracy difference relative to C2/C3; at $\alpha$ = 0.1, it costs roughly 1.2–1.3 percentage points of final accuracy because a larger fraction of benign updates also falls outside the acceptance region. This trade-off is operationally relevant in governance settings where an accepted contribution may trigger downstream credit, rewards, audit commitments, or eligibility decisions.

## 6.2 Heterogeneity as a Governance Constraint

The shift from $\alpha$ = 0.5 to $\alpha$ = 0.1 demonstrates why FL trust mechanisms cannot be designed as ordinary centralized anomaly filters. Under severe label skew, benign clients may legitimately disagree with the round reference and with a shared validation proxy. The peer-relative utility score mitigates this problem by judging a client against the current round rather than a fixed zero-gain boundary. Still, the lower benign retention rate shows the problem is not eliminated. The aggregate acceptance rates do not establish which benign clients were rejected or whether rejection disproportionately affects clients with highly skewed label distributions or smaller local datasets. Future trust policies should therefore adapt selectivity to observed heterogeneity, model convergence, or uncertainty, or use soft quarantine/probation states rather than a single permanent threshold. This observation is consistent with recent adaptive trust defenses that seek to separate malicious behavior from non-IID variation [8].

## 6.3 Connection to Blockchain-Assisted SDFL Governance

The present study evaluates only the off-chain trust scoring, screening, and aggregation mechanisms. These components can provide input to the broader blockchain-assisted SDFL framework introduced in [13], where screening decisions may support auditable participation and incentive management. The present results strengthen that architecture by replacing abstract trust inputs with reproducible update-level signals and by showing that trust status can separate benign and poisoning clients across multiple non-IID regimes. Importantly, the current paper does not claim blockchain performance, gas savings, incentive compatibility, or decentralized dispute resolution. Those remain separate system-layer questions. This separation also avoids attributing learning robustness to blockchain itself: the robustness observed here is produced by off-chain scoring, screening, and aggregation, while blockchain is best viewed as an audit and enforcement layer [10].

## 7 Limitations and Threats to Validity

The study has several limitations. The experiments use MNIST with a compact CNN, so generalization to higher-dimensional vision, language, or real cross-silo tasks remains untested. The adversarial evaluation is restricted to persistent 20% label-flipping clients; sign-flip, scaling, backdoor, free-rider, adaptive, colluding, and Sybil behaviors require separate study. Utility scoring relies on a 1,000-sample coordinator validation proxy $\mathcal{V}$, whose representativeness can affect trust decisions. The fixed $z_{i,t} \geq -3.5$ and $D_{i,t} \geq 0.55$ thresholds are not subjected to a full sensitivity sweep, and participant selection is intentionally random rather than trust-controlled. The admission, probation, rewards, stake-backed slashing, blockchain execution, and IPFS components of [13] are therefore outside the experimental scope. Finally, five seeds limit exact non-parametric test resolution, and one runtime measurement showed a clear environmental anomaly; runtime efficiency is consequently not a primary claim.

## 8 Conclusion

This paper operationalized and evaluated the trust-governance core of semi-decentralized FL under label-flipping attacks. Peer-relative utility normalization, similarity-based quality, dynamic trust, and trust-weighted robust aggregation retained near-baseline clean accuracy while rejecting 95.82% of malicious updates under moderate non-IID heterogeneity. Under severe heterogeneity, 88.55% were rejected, with lower benign retention and accuracy, exposing a utility-selectivity trade-off. Weighting-only methods preserved slightly higher utility while accepting all malicious submissions, confirming that influence mitigation and explicit containment are distinct objectives. Future work will extend the study to additional datasets, attacks, threshold sensitivity, fairness, trust-based admission, and the blockchain/IPFS enforcement layer.